\documentclass[11pt]{article}
\usepackage{amsmath}
\usepackage{amssymb}
\usepackage{amsthm} 
\usepackage{mathrsfs}
\usepackage[dvipdfmx]{graphicx,color}
\nopagebreak[3]

\newcommand{\newsection}[1]
{\section{#1}\setcounter{theorem}{0} \setcounter{equation}{0} \par\noindent}

\newtheorem{theorem}{Theorem}

\newtheorem{lemma}[theorem]{Lemma}
\newtheorem{corollary}[theorem]{Corollary}

\newtheorem{remark}[theorem]{Remark}

\newcommand{\beq}{ \begin{equation} }
\newcommand{\eeq}{ \end{equation} }

\newcommand{\bdot}{{\dot B}}
\newcommand{\hdot}{{\dot H}}
\newcommand{\br}{{\mathbb R}}

\title{
On the nonrelativistic limit of a semilinear field 
\\
equation in a uniform and isotropic space
}
\author
{
Makoto NAKAMURA
\thanks
{
{Faculty of Science, Yamagata University, 
Kojirakawa-machi 1-4-12, Yamagata 990-8560, JAPAN.}
E-mail: \texttt{nakamura@sci.kj.yamagata-u.ac.jp}
}
}
\date{}

\begin{document}

\maketitle

\begin{abstract}
The nonrelativistic limit of a semilinear field equation is considered in a uniform and isotropic space.
The scale-function of the space is constructed based on the Einstein equation.
The Cauchy problem of the limit-equation is considered, 
and global and blow-up solutions are shown in Sobolev spaces.
The role of spatial variance on the problem is studied, 
and some dissipative properties of the limit-equation are remarked.
\end{abstract}


%

\newsection{Introduction}
We consider a line element which has complex coefficients in a uniform and isotropic space.
We use the Einstein equation to set its scale-function which describes the spatial variance.
We consider a semilinear scalar-field equation in the space, 
and we take its nonrelativistic limit.
We consider the Cauchy problem of the limit-equation, 
and we show global and blow-up solutions in Sobolev spaces. 
We first introduce the limit-equation in this section, 
and we show its derivation in the next section.

We denote the spatial dimension by $n\ge1$, the Planck constant by $\hbar:=h/2\pi$, the mass by $m>0$.
Let $\sigma\in \mathbb{R}$, $a_0>0$, $a_1\in \mathbb{R}$.
We put $T_0:=\infty$ when $(1+\sigma)a_1\ge0$, 
$T_0:=-2a_0/n(1+\sigma)a_1(>0)$ when $(1+\sigma)a_1<0$.
We define a scale-function $a(t)$ for $t\in [0,T_0)$ by 
\begin{equation}
\label{a}
a(t):=
\left\{
\begin{array}{ll}
a_0\left( 1+\frac{n(1+\sigma)a_1t}{2a_0} \right)^{2/n(1+\sigma)}
& 
\mbox{if}\ \ \sigma \neq -1, 
\\
a_0\exp \left( \frac{a_1t}{a_0} \right) & 
\mbox{if}\ \ \sigma = -1,
\end{array}
\right.
\end{equation}
where we note that $a_0=a(0)$ and $a_1=\partial_t a(0)$.
We define a weight function 
$w(t):=(a_0/a(t))^{n/2}$,
and a change of variable $s=s(t):=\int_0^t a(\tau)^{-2}d\tau$.
We put $S_0:=s(T_0)$.
We use conventions $a(s):=a(t(s))$ and $w(s):=w(t(s))$ 
for $s\in [0,S_0)$ 
as far as there is no fear of confusion.
A direct computation shows 
\[
S_0=
\left\{
\begin{array}{ll}
\frac{2}{a_0 a_1(4-n(1+\sigma))}  & \mbox{if}\ \  
a_1\left(4-n(1+\sigma)\right)>0 
\\
\infty & \mbox{if}\ \  a_1\left(4-n(1+\sigma)\right)\le 0.
\end{array}
\right.
\]
For $\lambda \in \mathbb{C}$, $1\le p<\infty$, 
$-\pi/2<\omega\le \pi/2$, $0\le \mu_0< n/2$ and $0< S\le S_0$,  
we consider the Cauchy problem given by 
\beq
\label{Cauchy}
\left\{
\begin{array}{l}
\displaystyle
\pm i\frac{2m}{\hbar} \partial_s u(s,x)
+
e^{-2i\omega}\,\Delta u(s,x)
-
\lambda  e^{-2i\omega}a(s)^2 
\left(|uw|^{p-1}u\right)(s,x)=0, 
\\
u(0,\cdot)=u_0(\cdot)\in H^{\mu_0}(\br^n)
\end{array}
\right.
\eeq
for $(s,x)\in [0,S)\times\br^n$,
where $i:=\sqrt{-1}$,
$\Delta :=\sum_{j=1}^n \partial^2/\partial x_j^2$, 
and $H^{\mu_0}(\mathbb{R}^n)$ denotes the Sobolev space of order $\mu_0\ge0$.
The double sign $\pm$ is in same order throughout the paper.
We say that $u$ is a global solution of \eqref{Cauchy} if it exists on $[0,S_0)$.


When the scale-function is a constant $a(\cdot)=a_0=1$, 
we have $w(\cdot)=1$, $s=t$ and $S_0=T_0=\infty$.
The first equation in \eqref{Cauchy} corresponds to 
the Schr\"odinger equation 
\begin{equation}
\label{Schrodinger}
\pm i\frac{2m}{\hbar} \partial_s u(s,x)+\Delta u(s,x)
-
\lambda  \left(|u|^{p-1}u\right)(s,x)=0 
\end{equation}
when $\omega=0$,
and to the parabolic equation 
\begin{equation}
\label{Parabolic}
\frac{2m}{\hbar} \partial_s u(s,x)-\Delta u(s,x)
+
\lambda  \left(|u|^{p-1}u\right)(s,x)=0 
\end{equation}
when $\omega=\pm \pi/4$. 
The complex Ginzburg-Landau equation 
\begin{equation}
\label{CGL}
\frac{2m}{\hbar} \partial_s u(s,x)
-
\gamma\Delta u(s,x)
-
\lambda_1  u(s,x)
+
\lambda_2  \left(|u|^2u\right)(s,x)
=0 
\end{equation}
is also considered as the sum of the potentials with   
$p=1$ and $p=3$ 
for $0<\pm \omega< \pi/2$, 
where 
$\lambda_1\ge0$,  
$\gamma\in \mathbb{C}$ with $\mbox{\rm Re}\, \gamma>0$,
$\lambda_2\in \mathbb{C}$ with $\mbox{\rm Re}\, \lambda_2>0$.
We refer to 
\cite{Ginibre-Velo-1996-PhysicaD},
\cite{Li-Dai-Liu-2007-JMAA}, 
\cite[(2.1)]{Saarloos-Hohenberg-1992-PhysicaD}, 
\cite{Saarloos-Hohenberg-1993-PhysicaD}, 
\cite[(II.1) and the footnote in p.304]{Schmid-1966-PhysKondensMaterie} 
for the (generalized) complex Ginzburg-Landau equation.
We note that the dimension of $\hbar/m$ in the SI units is 
$\textsf{M}^2\textsf{S}^{-1}$ 
(\textsf{M}: meter, \textsf{S}: second),  
which is equivalent to the dimension of the thermal diffusivity 
$K_1$ of the heat equation 
$\partial_s u-K_1 \Delta u=0$, 
and also to the dimension of the diffusion coefficient $K_2$ of the diffusion equation 
$\partial_s u- K_2 \Delta u=0$.


\begin{remark}
Let us consider the diffusion equation in the space 
with a general scale-function $a(\cdot)>0$ 
which may be different from \eqref{a}.
We denote the diffusion constant by $K>0$.
For infinitesimal 
$\Delta t >0$ and $\Delta r>0$ 
with $(\Delta r)^2=2n K \Delta t$, 
we assume that the value of a function $\phi$ at $(t+\Delta t, x)$ is 
determined by the form  
\[
\phi(t+\Delta t,x)=
\frac{1}{n\omega_n}
\left(\frac{a(t)}{\Delta r}\right)^{n-1}
\int_{|y|=\Delta r/a(t)} 
\phi(t,x+y) d\sigma (y) 
+
\int_t^{t+\Delta t} f(\tau,x) d\tau,
\]
where $\omega_n$ denotes the volume of the unit ball in $\mathbb{R}^n$, 
$d\sigma$ denotes the measure on the sphere, 
and $f$ denotes a force.
By the mean value theorem and the divergence theorem, we have 
\[
\partial_t \phi(t,x)=\frac{ K }{a(t)^2} \Delta \phi(t,x)+f(t,x).
\]
We put $f:=-n(\partial_t a) \phi /2a -J V_1(\phi)$ for a constant 
$J$ and a function $V_1$.
We put the weight function $w(t):=(a(0)/a(t))^{n/2}$,
the change of variable $s=s(t):=\int_0^t a(\tau)^{-2} d\tau$, 
and a new function $u(t,x):=\phi(t,x)w(t)^{-1}$.
Then we have 
\[
\partial_s u(s,x)-K \Delta u(s,x)
+J a(s)^2 w(s)^{-1}
V_1(uw)(s, x)=0.
\]
This diffusion equation corresponds to the first equation in \eqref{Cauchy}, 
provided 
$\omega=\pm \pi/4$, 
$J=K=\hbar/2m$, 
and  $V_1(\phi)=\lambda  |\phi|^{p-1}\phi$.
We show a more general derivation of the first equation in \eqref{Cauchy}  starting from a line element in the next section.
\end{remark}

\begin{remark}
The first equation in \eqref{Cauchy} is rewritten as 
\[
\pm i\frac{2m}{\hbar} \partial_s u(s,x)
+
e^{-2i\omega}\,\Delta u(s,x)
-
\lambda  e^{-2i\omega}
a_0^{n(p-1)/2}a(s)^{2-n(p-1)/2} 
(|u|^{p-1}u)(s,x)=0.
\]
We have $a(\cdot)^{2-n(p-1)/2}=1$ 
when $p$ is the conformal power $p=1+4/n$.
This fact reminds us of the pseudo-conformal transform  
for \eqref{Schrodinger} 
(see \cite{Tsutsumi-Yajima-1984-BullAMS}).
Indeed,
the equation 
\[
\pm 2i \partial_t v(t,x)+e^{2i\omega}\Delta v(t,x)-\bar{\lambda} 
e^{2i\omega}(|v|^{p-1}v)(t,x)=0
\] 
is transformed to 
\[
\pm 2i \partial_s u(s,x)
+
e^{-2i\omega}\,\Delta u(s,x)
-
\lambda  e^{-2i\omega}
s^{n(p-1)/2-2}e^{\pm (p-1)(\sin 2\omega)x^2/2s} 
(|u|^{p-1}u)(s,x)=0
\]
by the transform  
$u(s,x)=s^{-n/2}e^{\pm ie^{2i\omega}x^2/2s} \bar{v} (1/s,x/s)$ 
with $t:=1/s$ and $y:=x/s$.
When $p=1+4/n$, $\omega=0$ and $\lambda\in \mathbb{R}$, 
the equation is invariant.
\end{remark}

It is well-known that the Schr\"odinger equation 
$i2m \partial_t u/\hbar+\Delta u-\lambda |u|^{p-1}u=0$ is derived from the Klein-Gordon equation 
$\partial_t^2 \phi -c^2\Delta \phi+(mc^2/\hbar)^2\phi
+c^2\lambda |\phi|^{p-1}\phi=0$ 
by the transform $\phi=u e^{-imc^2t/\hbar}$ and the nonrelativistic limit $c\rightarrow\infty$.
One of the aims of this paper is to show 
that the first equation in \eqref{Cauchy} 
which generalizes the Schr\"odinger equation and the parabolic equation is naturally derived 
from the nonrelativistic limit of the field equation \eqref{Eq-Field}, below. 
Moreover, we consider the spatial variance described by the scale-function $a(\cdot)$, which satisfies the Einstein equation with the cosmological constant 
in a uniform and isotropic space.
The study of roles of the cosmological constant 
and the spatial variance is important to describe the history of the universe, 
especially, the inflation and the accelerating expansion of the universe.  
The scale-function \eqref{a} follows from the equation of state when we regard the cosmological constant as the dark energy (see \eqref{Eq-State}, below).
In this paper, we study the cosmological constant from the point of view of 
partial differential equations.
We consider the Cauchy problem \eqref{Cauchy}, 
and we show the well-posedness of the problem, 
global solutions  
and blow-up solutions.
Especially, we remark that some dissipative properties appear by the spatial variance.


Let us consider the well-posedness of \eqref{Cauchy}.
For any real numbers $2\le q\le \infty$ and $2\le r<\infty$, 
we say that the pair $(q,r)$ is admissible if it satisfies 
$1/r+2/nq=1/2$.
For $\mu_0\ge0$ and two admissible pairs $\{(q_j,r_j)\}_{j=1,2}$, we define a function space 
\[
X^{\mu_0}([0,S)):=\{ u\in C([0,S), H^{\mu_0}(\mathbb{R}^n))
;\ \max_{\mu=0,\mu_0} \|u\|_{X^{\mu}([0,S))}<\infty\},
\]
where 
\[
\|u\|_{X^{\mu}([0,S))}
:=
\begin{cases}
\|u\|_{L^\infty((0,S), L^2(\mathbb{R}^n))
\cap
\bigcap_{j=1,2} L^{q_j} ((0,S), L^{r_j}(\mathbb{R}^n))}
& \mbox{if}\ \mu=0, \\
\|u\|_{L^\infty((0,S), \dot{H}^{\mu}(\mathbb{R}^n))
\cap
\bigcap_{j=1,2} L^{q_j} ((0,S),\bdot^{\mu}_{r_j2}(\mathbb{R}^n))}
& \mbox{if}\ \mu>0.
\end{cases}
\]
Here, $\hdot^{\mu}(\mathbb{R}^n)$ and 
$\bdot^{\mu}_{r_j2}(\mathbb{R}^n)$ denote the homogeneous Sobolev and Besov spaces, respectively.
Since the propagator of the linear part of the first equation 
in \eqref{Cauchy} is written as 
$\exp(\pm i\hbar \exp(-2i\omega) s\Delta/2m)$,
we assume $0\le \pm \omega\le \pi/2$ to define it 
as a pseudo-differential operator. 
We note that the scaling critical number of $p$ for \eqref{Cauchy} is 
$p(\mu_0):=1+4/(n-2\mu_0)$ when $a(\cdot)=1$.
We put 
\[
p_1(\mu_0):=1+\frac{4}{n-2\mu_0} \cdot 
\left(
1+\frac{4}{n-2\mu_0}\cdot \frac{2\mu_0}{n(1+\sigma)}
\right)^{-1} 
\]
for $\sigma\neq -1$.

\begin{theorem}
\label{Th-Power-LocalGlobal}
Let $n\ge1$, 
$\lambda \in \mathbb{C}$, 
$0\le \mu_0<n/2$, and $1\le p\le p(\mu_0)$.
Let $\omega$ satisfies $0\le \pm \omega\le \pi/2$ and $\omega\neq -\pi/2$.
Assume $\mu_0<p$ if $p$ is not an odd number. 
There exist two admissible pairs $\{(q_j,r_j)\}_{j=1,2}$ with the following properties.

(1) (Local solutions.) 
For any $u_0\in H^{\mu_0}(\mathbb{R}^n)$, 
there exist $S>0$ with $S\le S_0$ 
and a unique local solution $u$ of \eqref{Cauchy} in $X^{\mu_0}([0,S))$.
Here, $S$ depends only on the norm $\|u_0\|_{\hdot^{\mu_0}(\mathbb{R}^n)}$ when $p<p(\mu_0)$, while $S$ depends on the profile of $u_0$ when $p=p(\mu_0)$.
The solutions depend on the initial data continuously.

(2) (Small global solutions.) 
Assume that one of the following conditions from (i) to (vi) holds: 
(i) $\mu_0=0$, $p=p(0)$, 
(ii) $\mu_0>0$, $p=p(\mu_0)$, $a_1\ge0$,
(iii) $1<p<p(\mu_0)$, $a_1>0$, $\sigma<-1$, 
(iv) $1<p<p_1(\mu_0)$, $a_1<0$, $\sigma>-1$, 
(v) $p_1(\mu_0)<p<p(\mu_0)$, $a_1>0$, $\sigma>-1$, 
(vi) $\mu_0>0$, $1<p<p(\mu_0)$, $a_1>0$, $\sigma=-1$.
If $\|u_0\|_{\dot{H}^{\mu_0}(\mathbb{R}^n)}$ is sufficiently small, 
then the solution $u$ obtained in (1) is a global solution, namely, $S=S_0$. 
\end{theorem}


\begin{remark}
The results in Theorem \ref{Th-Power-LocalGlobal} also hold for the gauge variant equation  
\begin{equation}
\label{Eq-Schr-VPower-NonGauge}
\pm i\frac{2m}{\hbar} \partial_s u(s,x)
+e^{-2i\omega}\,\Delta u(s,x)
-\lambda e^{-2i\omega} \ \frac{a(s)^2}{w(s)} \ |uw|^p(s,x)=0,
\end{equation}
provided $\mu_0<p$ when $p$ is not an even number.
\end{remark}

\begin{remark}
The result (2) in Theorem \ref{Th-Power-LocalGlobal} especially shows 
that we always have small global solutions 
for $1<p<1+4/n$ when $a(\cdot)$ is not a constant in the conditions (iii) and (vi).
This result is much different from the case $a(\cdot)=1$ 
in the following sense.
In the case $a(\cdot)=1$ and $\omega=0$, 
namely \eqref{Schrodinger}, some weighted spaces, for example, $(1+|x|)^{-1}L^2(\mathbb{R}^n)$, have been needed for global solutions for  $1+2/n<p<1+4/n$ 
(see 
\cite{Ginibre-Ozawa-Velo-1994-AIHP,
Nakanishi-Ozawa-2002-NoDEA,
Tsutsumi-Yajima-1984-BullAMS}). 
There exist blow-up solutions for small initial data 
for $1<p\le 1+2/n$ 
(see \cite{Ikeda-Wakasugi-2013-DIE}). 
In the case $a(\cdot)=1$ and $\omega=\pm \pi/4$,  
namely \eqref{Parabolic}, 
there exist blow-up solutions for small initial data 
for $1<p\le 1+2/n$ 
(see 
\cite{Fujita-1966-JTokyo,
Hayakawa-1973-ProcJapanAcad,
Kavian-1987-AIHP,
Weissler-1981-IsraelJMath}).
\end{remark}


When $\lambda\in \mathbb{R}$, we are able to use the conservation law 
(see Lemma \ref{Lemma-Conservation}, below) 
to show global solutions for large data 
in $H^1(\mathbb{R}^n)$.
The global solutions were shown 
for the Schr\"odinger equation \eqref{Schrodinger} 
in \cite[Theorem 3.1]{Ginibre-Velo-1979-JFA-Part2}, 
for the complex Ginzburg-Landau equation \eqref{CGL} 
in \cite[Proposition 4.2]{Ginibre-Velo-1997-CMP}.
Blow-up solutions for initial data with negative energy are obtained  
by the concavity of an auxiliary function,  the virial identity, 
and the Heisenberg uncertainty principle.
We refer to 
\cite[Section 6.5]{Cazenave-2003-Courant} for \eqref{Schrodinger},
\cite[Theorem 1.8]{Banica-2007-CPDE} for \eqref{Schrodinger} in the hyperbolic space, 
and  
\cite[Theorem 5.3]{Hu-2011-LectureNotes} for \eqref{Parabolic}.
We show the following results for global and blow-up solutions for \eqref{Cauchy}.

\begin{corollary}
\label{Cor-GlobalEnergy-Power}
Let $\mu_0=0$ or $\mu_0=1$.
Let $\lambda >0$.
Let $1\le p<1+4/n$ when $\mu_0=0$. 
Let $1\le p<1+4/(n-2)$ and $a_1(p-1-4/n)\ge0$ when $\mu_0=1$.
For any $u_0\in H^{\mu_0}(\mathbb{R}^n)$, 
the local solution $u$ given by (1) in Theorem \ref{Th-Power-LocalGlobal} is a global solution.
\end{corollary}

\begin{corollary}
\label{Cor-GlobalEnergy-Power-Focusing}
Let $\mu_0=1$, $\lambda <0$, $a_1\ge0$ and $1\le p< 1+4/n$.
Let $\omega=0$ or $\omega=\pi/2$.
For any $u_0\in H^1(\mathbb{R}^n)$, 
the local solution $u$ given by (1) in Theorem \ref{Th-Power-LocalGlobal} is a global solution.
\end{corollary}

\begin{corollary}
\label{Cor-GlobalEnergy-Power-Focusing-BlowUp-Parabolic}
Let $\mu_0=1$ and $\lambda <0$. 
Let $\omega\neq0, \pi/2$.
Put $p_0:=2/(\sin 2\omega)^2-1$.
Let $p_0<p\le  1+4/(n-2)$.
Let $a_1(p-1-4/n)\le 0$ and $S_0=\infty$. 
For any $u_0\in H^1(\mathbb{R}^n)$ with negative energy 
\begin{equation}
\label{Eu0-Negative}
\int_{\mathbb{R}^n} \frac{1}{2}|\nabla u_0(x)|^2 
+ \frac{\lambda  a^2_0|u_0(x)|^{p+1}}{p+1} dx<0,
\end{equation}
the solution $u$ given by (1) in Theorem \ref{Th-Power-LocalGlobal} blows up in finite time.
\end{corollary}

\begin{corollary}
\label{Cor-GlobalEnergy-Power-Focusing-BlowUp-Dispersive}
Let $\mu_0=1$ and $\lambda <0$.
Let $\omega=0$ or $\omega=\pi/2$. 
Let $1+4/n\le p\le 1+4/(n-2)$.
Let $a_1\le 0$ and $S_0=\infty$.
For any $u_0\in H^1(\mathbb{R}^n)$ 
which satisfies $\||x|u_0(x)\|_{L^2_x(\mathbb{R}^n)}<\infty$ and \eqref{Eu0-Negative},
the solution $u$ given by (1) in Theorem \ref{Th-Power-LocalGlobal} blows up in finite time.
\end{corollary}

To prove the above corollaries, we use two dissipative properties.
One is from the parabolic structure of the first equation in \eqref{Cauchy} 
when $0<\pm \omega<\pi/2$.
The other is from the scale function $a(\cdot)$ when $\partial_ta(0)=a_1\neq 0$.
Even if the equation does not have the parabolic structure when 
$\omega=0, \pi/2$, 
the latter is very effective to obtain the global solutions.
This fact is expected 
from that the dissipative term $n\partial_t a\partial_t \phi/a$ 
in \eqref{Eq-Field-2}, below,  
appears as the weight function of $a(\cdot)$ and $w(\cdot)$ in \eqref{Cauchy}. 
For example, the energy estimate \eqref{Eus-Eu0}, below, 
shows the dissipative property of the equation 
when $\lambda  a_1(p-1-4/n)>0$ by \eqref{I-Rewrite}.
There is a large literature on the Cauchy problem for 
\eqref{Schrodinger} and \eqref{Parabolic} 
(see, for example,
\cite{
Cazenave-2003-Courant, 
Cazenave-Haraux-1998-Oxford, 
Cherrier-Milani-2012-AMS,
Tao-2006-AMS, 
Zheng-2004-Chapman}).
The properties of semilinear Schr\"odinger equations of the form 
$(i\partial_t+\Delta_g)u=|u|^{p-1}u$ 
have been studied on certain compact or noncompact Riemannian manifold $(M,g)$, 
where $\Delta_g$ is the Laplace-Beltrami operator on $(M,g)$.
In the hyperbolic space $\mathbb{H}^n$, 
the dispersive effect on Schr\"odinger equations was considered 
in \cite{Banica-2007-CPDE}, 
and the global existence of solutions with finite energy has been shown in 
\cite{Ionescu-Pausader-Staffilani-2012-AnalPDE}.
In the de Sitter spacetime,
a dissipative effect on Schr\"odinger equations was shown 
in \cite{Nakamura-2015-JDE}.

Throughout the paper, the notation $A\lesssim B$ denotes the inequality $A\le CB$ for some constant  $C>0$ which is not essential in our argument.
For any real number $1\le r\le \infty$, its conjugate number is denoted by $r'$ with $1/r+1/r'=1$.
For $\mu\in \mathbb{R}$ and $1\le r,m\le \infty$, we use 
the Lebesgue space $L^r(\mathbb{R}^n)$,
the Sobolev space $H^{\mu,r}(\mathbb{R}^n)$, 
the Besov space $B^{\mu}_{rm}(\mathbb{R}^n)$, 
and their homogeneous spaces 
$\hdot^{\mu,r}(\mathbb{R}^n)$ and $\bdot^{\mu}_{rm}(\mathbb{R}^n)$ 
(see \cite{Bergh-Lofstrom-1976-Springer} for the definitions and properties).


\newsection{Derivation of the equation}
We derive the scale-function $a(\cdot)$ in \eqref{a}, 
and the first equation in \eqref{Cauchy} in this section.
To do it, we generalize the Einstein equation and the Robertson-Walker metric 
for general dimensions and complex line elements 
(see \cite{Carroll-2004-Addison} and 
\cite{DInverno-1992-Oxford} 
for $n=3$ and real line elements). 
In the following, the Greek letters $\alpha, \beta,\gamma,\cdots$ run from $0$ to $n$, the Latin letters $j,k,\ell, \cdots$ run from $1$ to $n$.
We use the Einstein rule for the sum of indices of tensors, 
for example, 
${T^\alpha}_\alpha:=\sum_{\alpha=0}^n {T^\alpha}_\alpha$ 
and 
${T^i}_i:=\sum_{i=1}^n {T^i}_i$.
For the coordinate $x=(x^0,\cdots,x^n)$ in $\mathbb{R}^{1+n}$, we define the volume element 
$dV=\sum_{\sigma} \mbox{\rm sign} (\sigma) \, dx^{\sigma(0)} \cdots dx^{\sigma(n)}$, 
where $\sigma$ denotes the permutation of $\{0,\cdots, n\}$.
We consider a bilinear symmetric complex-valued functional $\langle \cdot, \cdot \rangle$ 
on the vector space spanned by the vectors 
$\{\partial/\partial x^\alpha\}_{\alpha=0}^n$. 
We put 
$g_{\alpha\beta}:=\langle \partial/\partial x^\alpha, \partial/\partial x^\beta\rangle$.
We denote by $(g_{\alpha\beta})$ the matrix 
whose components are given by $g_{\alpha\beta}$.
Put $g :={\rm det}(g_{\alpha\beta})$. 
Let $(g^{\alpha\beta})$ be the inverse matrix of 
$(g_{\alpha\beta})$.
We define the line element  
$g_{\alpha\beta} dx^\alpha dx^\beta$.
For the change of variables $y=y(x)$,  
we have 
$\sqrt{-g(y)} dV(y) 
= 
\left(\mbox{\rm sign}\, \mbox{det}(\partial y/\partial x)\right) 
\, \sqrt{-g(x)} dV(x)$, 
where  $g(y)$ denotes the determinant of 
$\left(g_{\alpha\beta}(y)\right)$ with 
$g_{\alpha\beta}(y):=\langle \partial/\partial y^\alpha, \partial/\partial y^\beta\rangle$,
where we take the square root of $-g$ as $-\pi< \mbox{arg} (-g)\le \pi$.
For any contravariant tensor $T^\alpha$, 
we denote its parallel displacement from $x$ to $x+dx$ by 
$\widetilde{T}^\alpha (x+dx)
:=T^\alpha(x)-{\Gamma^\alpha}_{\beta\gamma} (x)T^\beta(x) dx^\gamma$,
where ${\Gamma^\alpha}_{\beta\gamma}(x)$ 
denotes the proportional constant at $x$.
We assume that ${\Gamma^\alpha}_{\beta\gamma}$ satisfies the symmetry condition 
${\Gamma^\alpha}_{\beta\gamma}={\Gamma^\alpha}_{\gamma\beta}$ 
and 
\[
\left(g_{\alpha\beta}\widetilde{T}^\alpha\widetilde{T}^\beta\right)(x+dx)
=
\left(g_{\alpha\beta}{T}^\alpha {T}^\beta\right)(x)
+O\Big(\sum_{0\le \alpha\le n} (dx^\alpha)^2\Big)
\] 
for any $T^\alpha$ and $dx$. 
Then we have the Christoffel symbol
\[
{\Gamma^\alpha}_{\beta\gamma}
=
\frac{1}{2}
g^{\alpha\delta}
\left(
\partial_\beta g_{\delta\gamma}+\partial_\gamma g_{\beta\delta}
-\partial_\delta g_{\beta\gamma}
\right).
\]
We define the covariant derivative $\nabla_\beta$ 
for $T^\alpha$ by 
\begin{equation*}
\nabla_\beta T^\alpha
:=
\lim_{dx^\beta\rightarrow0} 
\frac{T^\alpha(x+\widehat{dx^\beta})-\widetilde{T}^\alpha(x+\widehat{dx^\beta})}{dx^\beta}
=\partial_\beta T^\alpha+{\Gamma^\alpha}_{\beta\gamma} T^\gamma,
\end{equation*}
where $\widehat{dx^\beta}:=(0,\cdots,0, dx^\beta,0,\cdots,0)$.
In general, we define  
\begin{multline*}
\nabla_\delta {T^{\alpha\beta \cdots} }_{\mu\nu\cdots}
:=
\partial_\delta {T^{\alpha\beta \cdots} }_{\mu\nu\cdots}
+
{\Gamma^\alpha}_{\delta\varepsilon} 
{T^{\varepsilon\beta \cdots} }_{\mu\nu\cdots}
+
{\Gamma^\beta}_{\delta\varepsilon} 
{T^{\alpha\varepsilon \cdots} }_{\mu\nu\cdots}
+\cdots
\\
-
{\Gamma^\varepsilon}_{\delta\mu} 
{T^{\alpha\beta \cdots} }_{\varepsilon\nu\cdots}
-
{\Gamma^\varepsilon}_{\delta\nu} 
{T^{\alpha\beta \cdots} }_{\mu\varepsilon\cdots}
-\cdots
\end{multline*}
for any tensor ${T^{\alpha\beta \cdots} }_{\mu\nu\cdots}$, 
the Riemann curvature tensor 
\[
{R^\delta}_{\alpha\beta\gamma}:=\partial_\beta {\Gamma^\delta}_{\alpha\gamma} 
-
\partial_\gamma {\Gamma^\delta}_{\alpha\beta}
+
{\Gamma^\delta}_{\varepsilon \beta} {\Gamma^\varepsilon}_{\alpha\gamma}
-
{\Gamma^\delta}_{\varepsilon\gamma} {\Gamma^\varepsilon}_{\alpha\beta}
\]
which is derived from ${R^\delta}_{\alpha\beta\gamma}T^\alpha
=(\nabla_\beta\nabla_\gamma-\nabla_\gamma\nabla_\beta)T^\delta$,
the Ricci tensor $R_{\alpha\beta}:={R^\gamma}_{\alpha\beta\gamma}$, 
the scalar curvature $R:=g^{\alpha\beta}R_{\alpha\beta}$.
We define the Einstein tensor by 
$G_{\alpha\beta}:=R_{\alpha\beta}-g_{\alpha\beta} R/2$ 
which satisfies the Euler-Lagrange equation $G_{\alpha\beta}=0$ 
for the Einstein-Hilbert action 
$\int_{\mathbb{R}^{1+n}}  R $ $\sqrt{-g} dx$.
The change of upper and lower indices is done by $g_{\alpha\beta}$ and $g^{\alpha\beta}$, 
for example, 
${G^\alpha}_\beta:=g^{\alpha\gamma}G_{\gamma\beta}$.

We put $r:=\left(\sum_{j=1}^n (x^j)^2\right)^{1/2}$ 
for any $(x^1,\cdots,x^n)\in \mathbb{R}^n$.
We assume that 
the space is uniform and isotropic,
and we consider the line element  
$(d\ell)^2:=-(dx^0)^2+e^{h(x^0)} e^{f(r)} \sum_{j=1}^n (dx^j)^2$, 
where $h$ is a real-valued function 
and  $f$ is a complex-valued function.
This line element is uniform in the sense that 
for any two points $P$ and $Q$ in $\mathbb{R}^n$, the ratio of the coefficients 
$e^{h(x^0)} e^{f(r_P)} /e^{h(x^0)} e^{f(r_Q)}$ is independent of $x^0$.
By a direct computation, we have ${G^0}_j={G^j}_0=0$, 
\[
{G^0}_0:=\frac{n-1}{2} 
\left(
\frac{n}{4}(\partial_0 h)^2-e^{-h-f}
\left( f''+(n-1)\frac{f'}{r}+\frac{n-2}{4} (f')^2 \right) 
\right),
\]
and 
\begin{multline*}
{G^j}_k:={g^j}_k 
\left(
\frac{n-1}{2}
\left(
\partial_0^2 h+\frac{n}{4}(\partial_0 h)^2
\right)
-\frac{n-2}{2}e^{-h-f}
\left(
f''+(n-2)\frac{f'}{r}+\frac{n-3}{4} (f')^2 
\right)
\right)
\\
+\frac{n-2}{2}
e^{-h-f} 
\left(
f''-\frac{f'}{r}-\frac{(f')^2}{2}
\right)
\frac{x^jx^k}{r^2},
\end{multline*}
where $f':=d f/d r$.
Since the space is isotropic,  the coefficient of $x^j x^k$ must vanish.
So that, we assume that $f$ satisfies $f''-f'/r-(f')^2/2=0$, by which we obtain 
\begin{equation}
\label{f}
e^f=q^2\left(1+\frac{k^2 r^2}{4} \right)^{-2}
\end{equation}
for constants $q(\neq0), k \in \mathbb{C}$.
For the stress-energy tensor 
${T^\alpha}_\beta:=\mbox{\rm diag} (\rho c^2, -p,\cdots, -p)$ 
of the perfect fluid with the density $\rho$ and the pressure $p$, 
we define the $(1+n)$-dimensional Einstein equation 
\begin{equation}
\label{Einstein}
{G^\alpha}_{\beta}-\Lambda {g^\alpha}_\beta
= \kappa \, \, {T^\alpha}_\beta, 
\end{equation}
where 
$\Lambda\in \mathbb{R}$ is the cosmological constant, 
and the constant $\kappa$ is a positive real constant.
We assume $\kappa c^4$ is a constant independent of $c$
(see Remark \ref{Remark-Kappa}).

\begin{remark}
\label{Remark-Kappa}
When $n=3$, $q,k\in \mathbb{R}$, the constant $\kappa$ is called 
the Einstein gravitational constant which is given by 
$\kappa=8\pi \mathcal{G}/c^4$, 
where $\mathcal{G}$ is the Newton gravitational constant.
For $n\ge 3$, $q,k\in \mathbb{C}$, the constant $\kappa$ is generalized to 
\begin{equation}
\label{Kappa}
\kappa:=\frac{2(n-1)\pi^{n/2} \mathcal{G}}{(n-2)\Gamma(n/2) c^{4} }.
\end{equation}
Indeed, we define the equation of motion in the $(1+n)$-dimensional spacetime by 
$d^2 x^j/dt^2+\partial_j \phi=0$,
where $\phi:=-\mathcal{G} \rho \ast |x|^{2-n}/(n-2)$ is the gravitational potential and $\rho$ denotes the density function of mass.
We denote the volume of the unit ball in $\mathbb{R}^n$ by 
$\omega_n:=2\pi^{n/2}/n\Gamma(n/2)$.
The potential $\rho$ satisfies the Poisson equation 
$\Delta \phi=n\omega_n \mathcal{G}\rho$.
The Euler-Lagrange equation for the Lagrangian 
$\left\{-g_{\alpha\beta} (dx^\alpha/d\tau)(dx^\beta/d\tau)\right\}^{1/2}$ 
yields the geodesic equation 
$d^2x^\gamma/d\tau^2+{\Gamma^\gamma}_{\alpha\beta}
(dx^\alpha/d\tau)(dx^\beta/d\tau)=0$,
where $d\tau$ is the proper time defined by $(d\tau)^2=-(d\ell)^2$.
We consider a line element 
$\eta_{\alpha\beta} dx^\alpha dx^\beta$ 
defined by  $(\eta_{\alpha\beta})=\mbox{\rm diag}(-1, C,\cdots,C)$ 
for a complex number $C$.
We assume that $g_{\alpha\beta}$ is stationary 
$\partial_0 g_{\alpha\beta} =0$ 
and symmetric $g_{\alpha\beta}=g_{\beta\alpha}$, 
and 
$g_{\alpha\beta}$ is weak in the sense that 
$|g_{\alpha\beta}-\eta_{\alpha\beta}|$ is sufficiently small.
Moreover, we assume that the velocity of particles is 
very slow compared with the light speed $c$, 
and the pressure is almost zero, 
namely, $|dx^j/cdt|\ll 0$ and $p\ll1$.
Then we have 
${\Gamma^0}_{00}=0$, ${\Gamma^j}_{00}\doteqdot-\partial_j g_{00}/2C$, 
and 
$T_{\alpha\beta}/c^2\doteqdot 
\mbox{\rm diag} (-\rho,0,\cdots,0)$ 
by the definition of ${\Gamma^\alpha}_{\beta\gamma}$ and $T_{\alpha\beta}$. 
We have 
$R_{00}\doteqdot -\partial_j {\Gamma^j}_{00}$ by the definition of the Ricci tensor.
We have $(d^2 x^\alpha/dt^2)/c^2 +{\Gamma^\alpha}_{00}\doteqdot 0$  
by the geodesic equation, which yields 
$\partial_j(\phi/c^2+ g_{00}/2C)\doteqdot 0$ 
by the equation of motion and ${\Gamma^j}_{00}\doteqdot -\partial_j g_{00}/2C$.
So that, we have 
$R_{00}\doteqdot \Delta g_{00}/2C \doteqdot -\Delta\phi/c^2
=-n\omega_n \mathcal{G} \rho/c^2$.
Since the Einstein equation \eqref{Einstein} with $\Lambda=0$ is rewritten as 
$R_{\alpha\beta}=\kappa (T_{\alpha\beta}-g_{\alpha\beta} T/(n-1))$, 
we obtain \eqref{Kappa} from 
$R_{00}=\kappa (T_{00}-g_{00} T/(n-1))$, 
$T_{00}\doteqdot-\rho c^2$ and $T\doteqdot \rho c^2$. 
\end{remark}

We define a positive function 
$b(x^0):=e^{h(x^0)/2}$.
We put $\widetilde{\rho}:=\rho+\Lambda /\kappa c^2$, and 
$\widetilde{p}:=p-\Lambda/\kappa$.
Then \eqref{Einstein} is rewritten as  
${G^\alpha}_\beta
=
\kappa \cdot \mbox{diag}(\widetilde{\rho}c^2, -\widetilde{p}, \cdots, -\widetilde{p})$, 
which shows that the cosmological constant $\Lambda>0$ is regarded 
as the energy which has positive density and negative pressure in the vacuum 
$\rho=p=0$ 
(``the dark energy" for $n=3$). 
The equation ${G^0}_{0}=\kappa \widetilde{\rho} c^2{g^0}_0$ is equivalent to 
\begin{equation}
\label{G00}
\frac{n-1}{2}
\left\{
\left(
\frac{\partial_0 b}{b}
\right)^2
+
\frac{k^2}{q^2b^2}
\right\}
=\frac{\kappa c^2}{n} \cdot \widetilde{\rho}. 
\end{equation}
The equation ${G^j}_{k}=-\kappa \widetilde{p} {g^j}_k$ is equivalent to 
\begin{equation}
\label{Gjk}
\frac{n-1}{2}
\left\{
\frac{2}{n-2} \cdot \frac{\partial_0^2b}{b}
+
\left(
\frac{\partial_0 b}{b}
\right)^2
+
\frac{k^2}{q^2b^2}
\right\}
=-\frac{\kappa}{n-2}\cdot  \widetilde{p},
\end{equation}
which is rewritten as the Raychaudhuri equation 
\begin{equation}
\label{Ray}
\frac{\partial_0^2 b}{b}
=-\frac{n-2}{n-1}\cdot \kappa 
\left(
\frac{\widetilde{\rho}c^2}{n}
+
\frac{\widetilde{p}}{n-2}
\right)
\end{equation}
by \eqref{G00}.
Multiplying $b^n$ to the both sides in \eqref{G00}, 
and taking the derivative by $x^0$ variable, 
we have the conservation of mass  
\begin{equation}
\label{Conservation-Mass}
\partial_0(\widetilde{\rho} c^2 b^n)+\widetilde{p} \partial_0 b^n=0.
\end{equation} 
For any real number $\sigma$, we assume the equation of state 
\begin{equation}
\label{Eq-State}
\widetilde{p}=\sigma \widetilde{\rho} c^2.
\end{equation}
Then $b(x^0)$ must satisfy 
\[
\frac{\partial_0^2 b(x^0)}{b(x^0)}
=-\frac{2}{n-1}\cdot \frac{n-2+n\sigma}{2n} \cdot \kappa \widetilde{\rho} c^2 
\ \ \ \ 
\mbox{with}
\ \ \ \ 
\widetilde{\rho}=\frac{n-1}{2}\cdot \frac{n}{\kappa c^2}
\cdot \frac{\partial_0 b(0)^2}{b(0)^{2-n(1+\sigma)} } 
\cdot b(x^0)^{-n(1+\sigma)}
\]
by \eqref{Ray} and \eqref{Conservation-Mass}, which has the solution 
\begin{equation}
\label{b}
b(x^0):= 
\left\{
\begin{array}{ll}
b(0)\left( 1+\frac{n(1+\sigma)\partial_0b(0)}{2b(0)} x^0\right)^{2/n(1+\sigma)}
& 
\mbox{if}\ \ \sigma \neq -1, 
\\
b(0)\exp \left( \frac{\partial_0b(0) x^0}{b(0)} \right) & 
\mbox{if}\ \ \sigma = -1.
\end{array}
\right.
\end{equation}
By \eqref{G00} and \eqref{b}, we have $k=0$.

For any $\lambda_1 \in \mathbb{C}$ and any complex-valued $C^2$ function $\phi$ on $\mathbb{R}^{1+n}$, we define the Lagrangian 
\[
L(\phi):=-\frac{1}{2} g^{\alpha\beta} \partial_\alpha \phi \, \partial_\beta \phi
-
\frac{1}{2}\left(\frac{mc}{\hbar}\right)^2\phi^2
+
\frac{\lambda_1 }{p+1}|\phi|^{p-1}\phi^2,
\]
and we apply the variational method to the action  
$\int_{\mathbb{R}^{1+n}} L(\phi) \sqrt{-g} dx$ for $\phi$. 
Then the Euler-Lagrange equation is given by 
\begin{equation}
\label{Eq-Field}
\frac{1}{ \sqrt{-g} } \partial_\alpha 
(\sqrt{-g} g^{\alpha\beta} \partial_\beta \phi)
-
\left(\frac{mc}{\hbar}\right)^2\phi 
+
\lambda_1  |\phi|^{p-1}\phi =0
\end{equation}
under the constraint condition $\mbox{arg}\, \delta \phi= \mbox{arg}\, \phi$.
This is the equation of motion of massive scalar field described by a function $\phi$ with the mass $m$ and the potential $\lambda_1  |\phi|^{p-1}\phi^2/(p+1)$.

We put $x^0:=ct$.
Then we have $a(t)=b(x^0)$ by \eqref{a} and \eqref{b}.
We put $q=e^{i\omega}$ for $-\pi/2<\omega\le \pi/2$ in \eqref{f}.
Then the line element $d\ell$ is rewritten as 
$d\ell^2=-(cdt)^2+a(t)^2e^{i2\omega}\sum_{j=1}^n (dx^j)^2$.
Then the field equation \eqref{Eq-Field} is rewritten as 
\begin{equation}
\label{Eq-Field-2}
-\frac{1}{c^2}\left(
\partial_t^2+\frac{n\partial_t a}{a}\partial_t+\frac{m^2c^4}{\hbar^2}
\right)\phi
+
\frac{1}{a^{2}e^{i2\omega}}  \Delta \phi
+
\lambda_1  |\phi|^{p-1}\phi=0.
\end{equation}
We refer to \cite{Galstian-Yagdjian-2015-NonlinearAnal}, \cite{Nakamura-2014-JMAA} and the references therein 
on energy solutions of  \eqref{Eq-Field-2} when $\omega=0$.
We define a function $u$ by 
\[
u(t, x^1, \cdots, x^n) :=\phi(x)
\left(\frac{a(t)}{a_0}\right)^{n/2} \exp\left(\pm i \frac{m}{\hbar}c^2 t \right).
\]
Then the nonrelativistic limit ($c\rightarrow\infty$) of \eqref{Eq-Field-2} yields 
\[
\pm i\frac{2m}{\hbar} \partial_t u(t,x)+a(t)^{-2} e^{-2i\omega} \Delta u(t,x)
+ 
\lambda_1 \left(|u w|^{p-1} u\right)(t,x)=0.
\]
By the change of variable $s:=\int_0^t a(\tau)^{-2}d\tau$, we obtain the first equation in \eqref{Cauchy} as $\lambda_1 :=-\lambda e^{-2i\omega}$.
Since $u=u(t,\cdot)$ is a global solution if it exists on $[0,T_0)$, 
we say $u=u(s,\cdot)=u(t(s),\cdot)$ is a  global solution 
if it exists on $[0,S_0)$.

%

\begin{remark}
In the above argument, we have derived the line element of the form 
\[
-(cdt)^2+
a(t)^2q^2\left(1+\frac{k^2 r^2}{4} \right)^{-2}\sum_{j=1}^n (dx^j)^2
\]
for constants $q(\neq 0), k \in \mathbb{C}$ 
as the solution of \eqref{Einstein}.
This is known as the Robertson-Walker metric 
when $n=3$, $q=1$ and $k^2=0, \pm 1$. 
Here, $k^2$ denotes the curvature of the space.
We note that $a(\cdot)$ in \eqref{a} blows up in finite time 
when $a_1>0$ and $\sigma<-1$, 
which is called Big-Rip in cosmology. 
The case $\sigma= -1$ shows the exponential expansion 
of $a(\cdot)$ when $a_1>0$.
The case $\sigma> -1$ shows the polynomial expansion 
of $a(\cdot)$ when $a_1>0$.
These models are studied for the accelerating expansion  
of the universe when $n=3$.
\end{remark}

%

\begin{remark}
We have assumed the equation of state 
\eqref{Eq-State} 
to derive \eqref{b}.
Instead of this equation, let us assume 
$p=0$, $\rho\ge0$, $\Lambda\in \mathbb{R}$, 
$k^2/q^2\in \mathbb{R}$.
Then we have the $(1+n)$-dimensional version of the $(1+3)$-dimensional models of the universe as follows.
By \eqref{G00},  
$b(\cdot)$ must satisfy 
\[
(\partial_0 b)^2=Rb^{2-n}+Lb^2-\frac{k^2}{q^2},
\ \ 
R:=\frac{2\kappa \rho c^2 b^n}{n(n-1)},
\ \ 
L:=\frac{2\Lambda}{n(n-1)},
\]
where $R$ is a constant by the conservation of mass.
We have the following solutions for this differential equation.
(1) If $\rho=\Lambda=0$, then $k^2/q^2\le 0$ and 
$b(x^0)=b(0)\pm \sqrt{-k^2/q^2} x^0$.
These solutions are the Minkowski model ($k=0$), 
and the Milne model ($k^2/q^2<0$) when $n=3$.
(2) If $\rho=0$ and $\Lambda<0$, then $k^2/q^2\le 0$ and 
$b(x^0)=|k^2/q^2 L|^{1/2} \sin (|L|^{1/2} x^0)$.
(3) If $\rho=0$ and $\Lambda>0$, then  
$b(x^0)=b(0)e^{\pm L^{1/2}x^0}$ when $k^2/q^2=0$, 
$b(x^0)=(k^2/q^2 L)^{1/2} \cosh (L^{1/2} x^0)$ when $k^2/q^2>0$,
$b(x^0)=|k^2/q^2 L|^{1/2} \sinh (L^{1/2} x^0)$ when $k^2/q^2<0$.
These solutions are the de Sitter models when $n=3$.
(4) 
If $\rho>0$ and $\Lambda=0$, then  
$b(x^0)=(b(0)^{n/2}\pm nR^{1/2} x^0/2)^{2/n}$ when $k^2/q^2=0$ 
(Einstein-de Sitter model for $n=3$), 
$b(x^0)=(q^2R (1-\cos2\theta)/2k^2)^{1/(n-2)}$ 
with 
$\int_0^\theta ((1-\cos2\theta)/2)^{1/(n-2)} d\theta
=
(k^2/q^2)^{1/2+1/(n-2)} R^{-1/(n-2)} (n-2)x^0/2$ 
when $k^2/q^2>0$, 
$b(x^0)=(|q^2|R (\cosh2\theta-1)/2|k^2|)^{1/(n-2)}$ 
with 
$\int_0^\theta ((\cosh2\theta-1)/2)^{1/(n-2)} d\theta$
$=|k^2/q^2|^{1/2+1/(n-2)} $ 
$R^{-1/(n-2)} (n-2)x^0/2$ 
when $k^2/q^2<0$.
These solutions are the Friedmann model when $n=3$.
(5) 
If $\rho>0$ and $\Lambda>0$, then  
$b(x^0)$ tends to infinity as $x^0$ tends to infinity 
for the cases 
(i) $k^2/q^2\le 0$,
or 
(ii) $k^2/q^2>0$ and $L>L_0$ (the Lema\^{i}tre model when $n=3$), 
or 
(iii) $k^2/q^2>0$ with $L= L_0:=(k^2/2q^2)^{n/(n-2)} R^{-2/(n-2)}$ 
and  $b(0)>b_\ast:=(R/L)^{1/n}$ (the Eddington-Lema\^{i}tre model when $n=3$).
For the case $k^2/q^2>0$, $L=L_0$ and $b(0)=b_\ast$, we have $b(\cdot)=b_\ast$
 (the Einstein model when $n=3$).
For the case $k^2/q^2>0$, $L=L_0$ and $b(0)<b_\ast$, $b(\cdot)$ tends to $b_\ast$ from below.
For the case $k^2/q^2>0$, $L<L_0$ and $b(0)<b_\ast$, 
$b(\cdot)$ is bounded by $b_\ast$ from above and vanishes in finite time.
For the case $k^2/q^2>0$, $L<L_0$ and $b(0)>b_\ast$, 
$b(\cdot)$ is bounded by $b_\ast$ from below and tends to infinity 
as $x^0$ tends to infinity.
\end{remark}

%

\begin{remark}
Let us consider the relation between the line element 
$-(c dt)^2+a(t)^2e^{i2\omega}$ $\sum_{j=1}^n (d x^j)^2$ 
and transforms in spacetime. 
Let us consider two coordinates 
$x=(ct, x^1,\cdots, x^n)$ and 
$x_\ast=(ct_\ast, x_\ast^1,$ $ \cdots, x_\ast^n)$.
We assume that the coordinate $x_\ast$ moves in the coordinate $x$ 
to infinity along $x^1$ axis 
at velocity $v\in \mathbb{R}$, 
and the origins in the coordinates coincide when $t=t_\ast=0$.
By the spatial rotation in $x_\ast$, 
we may assume $x^j=x_\ast^j$ for $2\le j\le n$. 
So that, it suffices to consider the relation between 
$(ct,x^1)$ and $(ct_\ast,x_\ast^1)$.
Since we consider the uniform linear motion, 
$(ct,x^1)$ and $(ct_\ast,x_\ast^1)$ has the transform 
$ ^T(ct_\ast,x_\ast^1)=A(v)\, ^T(ct,x^1)$ 
by a matrix 
$A(v):=
\begin{pmatrix}
a_{00} & a_{01} \\
a_{10} & a_{11}
\end{pmatrix}$ 
which is dependent on $v$,
where $ ^T(ct,x^1)$ denotes the transpose of $(ct,x^1)$, 
and $a_{00},a_{01},a_{10},a_{11}\in \mathbb{R}$. 
Since the point $(ct_\ast,0)$ in $x_\ast$-coordinate is observed 
as $(ct, vt)$ in $x$-coordinate, 
and the point $(ct, 0)$ in $x$-coordinate is observed 
as $(ct_\ast, -vt_\ast)$ in $x_\ast$-coordinate,
we have $a_{10}=-a_{11}v/c$ and $a_{11}=a_{00}$.
Since the coordinate 
$(ct,x^1)$ moves at velocity $-v$ in the coordinate $(ct_\ast,x_\ast^1)$ 
by the relativity, 
which means that 
$(ct,-x^1)$ moves at velocity $v$ in $(ct_\ast,-x_\ast^1)$, 
we have 
$ ^T(ct,-x^1)=A(v) ^T(ct_\ast,-x_\ast^1)$.
So that, we have $a_{00}^2-a_{01}a_{10}=1$ 
since   
$\begin{pmatrix}
ct_\ast \\
-x_\ast^1
\end{pmatrix}
=
\begin{pmatrix}
a_{00} & a_{01} \\
-a_{10} & -a_{11}
\end{pmatrix}
\begin{pmatrix}
ct \\
x^1
\end{pmatrix}$.
Therefore, the components of $A(v)$ must satisfy  
\begin{equation}
\label{Av}
a_{11}=a_{00},\ \ a_{10}=-\frac{v}{c} \cdot a_{00},\ \ a_{00}^2-a_{01}a_{10}=1.
\end{equation}
Now, we assume the invariance of the line element, namely, 
\begin{equation}
\label{Line-Element}
-(cdt_\ast)^2+a(t_\ast)^2e^{i2\omega}\sum_{j=1}^n (dx_\ast^j)^2 
=
-(cdt)^2+a(t)^2e^{i2\omega}\sum_{j=1}^n (dx^j)^2 
\end{equation} 
When $a(\cdot)=0$, we have $cdt_\ast=cdt$.
Then we have $a_{00}=1$ and the Galilei transform 
$A(v)=
\begin{pmatrix}
1 & 0 \\
-v/c & 1
\end{pmatrix}$ 
by \eqref{Av}.
When $a(\cdot)=1$ and $\omega=\pi/2$, 
we have 
$a_{00}^2+a_{10}^2=1$ and $a_{01}=-a_{10}$ by \eqref{Line-Element}.
Then we have the rotation of the spacetime
$A(v)=
(1+(v/c)^2)^{-1/2}
\begin{pmatrix}
1 & v/c \\
-v/c & 1
\end{pmatrix}$ by \eqref{Av}.
When $a(\cdot)=1$ and $\omega=0$, 
we have 
$a_{00}^2-a_{10}^2=1$ and $a_{01}=a_{10}$ by \eqref{Line-Element}.
Then we have the Lorentz transform 
$A(v)=
(1-(v/c)^2)^{-1/2}
\begin{pmatrix}
1 & -v/c \\
-v/c & 1
\end{pmatrix}$ 
by \eqref{Av}.
When $a(\cdot)=1$ and $\omega\neq0, \pi/2$, 
we have $cdt_\ast=cdt$, $dx^1=dx_\ast^1$, 
$a_{00}=a_{11}=1$ and $a_{01}=a_{10}=0$ by \eqref{Line-Element}, 
which requires $v=0$ by \eqref{Av}.
Therefore, the rotational transform for the spatial variables 
$(x^1, \cdots, x^n)$ is only allowed in the case 
$\omega\neq0,\pi/2$.
\end{remark}

%

\begin{remark}
We note that the line element with complex coefficients naturally unifies the rotational transform and the Lorentz transform as follows.
Let us consider two coordinates 
$z=(z^0,z^1, \cdots, z^n)\in \mathbb{C}^{1+n}$ and 
$z_\ast=(z_\ast^0,z_\ast^1, \cdots, z_\ast^n)\in \mathbb{C}^{1+n}$ which satisfy the invariance of the line elements 
$\sum_{\alpha=0}^n (dz^\alpha)^2=\sum_{\alpha=0}^n (dz_\ast^\alpha)^2$.
We assume 
$z^j=z_\ast^j$ for $2\le j\le n$.
For $\theta\in \mathbb{C}$, 
the transform  
$\begin{pmatrix}
z_\ast^0 \\
z_\ast^1
\end{pmatrix}
= 
\begin{pmatrix}
\cos\theta & -\sin\theta \\
\sin\theta &\cos\theta
\end{pmatrix}
\begin{pmatrix}
z^0 \\
z^1
\end{pmatrix}$ 
satisfies this invariance.
For any fixed $-\pi/2<\omega\le \pi/2$, 
let us consider the lines 
$z^0=ict$, $z^1=x^1e^{i\omega}$,
$z_\ast^0=ict_\ast$ and $z_\ast^1=x_\ast^1 e^{i\omega}$ 
in the complex plane $\mathbb{C}$,
where $t,t_\ast, x^1,x_\ast^1\in \mathbb{R}$.
Then we have the transform 
$\begin{pmatrix}
ct_\ast \\
x_\ast^1
\end{pmatrix}
=
\begin{pmatrix}
\cos\theta & i e^{i\omega}\sin\theta \\
i e^{-i\omega} \sin\theta &\cos\theta
\end{pmatrix}
\begin{pmatrix}
ct \\
x^1
\end{pmatrix}$.
So that, 
if $\omega=\pi/2$ and $\theta  \in \mathbb{R}$, 
then we have the rotational transform 
$\begin{pmatrix}
ct_\ast \\
x_\ast^1
\end{pmatrix}
=
\begin{pmatrix}
\cos\theta & -\sin\theta \\
\sin\theta &\cos\theta
\end{pmatrix}
\begin{pmatrix}
ct \\
x^1
\end{pmatrix}$.
If $\omega=0$ and $i\theta \in \mathbb{R}$, 
then we have the Lorentz transform  
$\begin{pmatrix}
ct_\ast \\
x_\ast^1
\end{pmatrix}
=
\begin{pmatrix}
\cosh(i\theta) & \sinh(i \theta) \\
\sinh(i\theta) & \cosh(i\theta)
\end{pmatrix}
\begin{pmatrix}
ct \\
x^1
\end{pmatrix}$.
\end{remark}

%

\newsection{Proof of Theorem \ref{Th-Power-LocalGlobal} }
We define the operator $U_{\pm}(s):=\exp(\pm i \hbar \exp(-2i\omega) s\Delta /2m)$.
We put $V_1(\phi):=\lambda |\phi|^{p-1}\phi$ and 
$f(s,u):=-i\hbar e^{-2i\omega} a(s)^2 w(s)^{-1}V_1((uw)(s))/2m$. 
We regard the solution of the Cauchy problem of \eqref{Cauchy} as the fixed point of the integral equation given by 
\begin{equation}
\label{Map-Phi}
u(s)=u(s,\cdot)=\Phi(u)(s):=U_{\pm}(s)u_0\pm \int_0^s U_\pm(s-\tau)f(\tau, u(\tau))d\tau.
\end{equation}
We define $\{(q_j,r_j)\}_{j=0}^2$ by $1/r_j=1/2-(p-1)(n-2\mu_0)/(p+1)2n$ and $1/r_j+2/nq_j=1/2$ for $j=0,1,2$.
Then $\{(q_j,r_j)\}_{j=0}^2$ are admissible pairs by the condition on $p$. 
We define $q(\mu)$ by 
$1/q(\mu)=1-(p-1)(n-2\mu)/4$  
for $0\le \mu<n/2$, 
and $r_\ast$ by 
$1/r_\ast:=1/r_1-\mu_0/n$.
Then we have 
$1/q(\mu_0):=1/q_0'-(p-1)/q_1-1/q_2$,  
$r_\ast<\infty$ 
and 
$1/r_0'=(p-1)/r_\ast+1/r_2$.
We have 
\begin{equation*}
\|\Phi(u)\|_{X^0([0,S))}
\lesssim 
\|u_0\|_{L^2(\mathbb{R}^n)}
+\|f(\cdot, u)\|_{L^{q_0'}((0,S),L^{r_0'}(\mathbb{R}^n)) } 
\end{equation*}
and 
\begin{equation}
\label{Strichartz-Phi}
\|\Phi(u)\|_{X^\mu([0,S))}
\lesssim 
\|u_0\|_{\hdot^\mu(\mathbb{R}^n)}
+\|f(\cdot, u)\|_{L^{q_0'}((0,S),\bdot^\mu_{r_0'2}(\mathbb{R}^n)) }
\end{equation}
for any $\mu\in \mathbb{R}$ 
by the condition $0\le \pm \omega\le \pi/2$, 
the energy estimate for $\omega\neq 0$ and $\omega\neq \pi/2$ 
(see, for example, \cite[Lemma 2.1]{Nakamura-2011-RMP}), 
and the Strichartz estimate for $\omega =0$ and $\omega=\pi/2$ 
(see \cite{Cazenave-2003-Courant}, \cite{Ginibre-Velo-1995-JFA}).
By Lemma 2.2 in \cite{Nakamura-Ozawa-1997-RMP}, we have 
\[
\|f(\cdot,u)\|_{\bdot^\mu_{r_0'2}(\mathbb{R}^n) }
\lesssim 
\frac{\hbar|\lambda|}{m}  a^2 w^{p-1} 
\|u\|_{ L^{r_\ast}(\mathbb{R}^n)\cap \bdot^{0}_{r_\ast 2}(\mathbb{R}^n) }^{p-1} 
\|u\|_{\bdot^\mu_{r_22}(\mathbb{R}^n) },
\]
where the assumption $\mu<p$ is required if $p$ is not an odd number.
By the embedding 
$\bdot^{\mu_0}_{r_12}(\mathbb{R}^n)
\hookrightarrow 
L^{r_\ast}(\mathbb{R}^n) 
\cap 
\bdot^0_{r_\ast 2}(\mathbb{R}^n)$ 
and the H{\"o}lder inequality in time, we obtain 
\[
\|f(\cdot,u)\|_{L^{q_0'}((0,S),\bdot^\mu_{r_0'2}(\mathbb{R}^n)) }
\lesssim 
\frac{\hbar|\lambda|}{m}A 
\|u\|_{L^{q_1} ((0,S), \bdot^{\mu_0}_{r_12}(\mathbb{R}^n)) }^{p-1} 
\|u\|_{L^{q_2} ((0,S), \bdot^{\mu}_{r_22}(\mathbb{R}^n)) },
\]
where we have put 
$A:=\|a(s)^2w(s)^{p-1}\|_{L_s^{q(\mu_0)}((0,S))}$.
Analogously, we also have 
\[
\|f(\cdot,u)\|_{L^{q_0'}((0,S),L^{r_0'}(\mathbb{R}^n)) }
\lesssim 
\frac{\hbar|\lambda|}{m}A 
\|u\|_{L^{q_1} ((0,S), \bdot^{\mu_0}_{r_12}(\mathbb{R}^n)) }^{p-1} 
\|u\|_{L^{q_2} ((0,S), L^{r_2}(\mathbb{R}^n)) }
\]
and 
\begin{eqnarray}
d(\Phi(u),\Phi(v))
&\lesssim& 
\|f(\cdot, u)-f(\cdot,v)\|_{L^{q_0'}((0,S), L^{r_0'}(\mathbb{R}^n)) }
\nonumber \\
&\lesssim& 
\frac{\hbar|\lambda|}{m}A \max_{w=u,v} 
\|w\|_{L^{q_1} ((0,S), \bdot^{\mu_0}_{r_12}(\mathbb{R}^n)) }^{p-1} 
d(u,v).
\label{Es-Metric}
\end{eqnarray}
So that, $\Phi$ is a contraction mapping on the ball  
\[
X(R_0,R_{\mu_0}):=\{u\in 
X^{\mu_0}([0,S)) 
\, ;\, \|u\|_{X^\mu([0,S))}\le R_\mu\ \mbox{for}\ \mu=0,\mu_0\}
\]
with the metric $d$ if $R_0$ and $R_{\mu_0}$ satisfy 
\begin{equation}
\label{Condition-Contraction}
R_0  \ge 2C_0\|u_0\|_{ L^2(\mathbb{R}^n) }, 
\ \ 
R_{\mu_0}  \ge 2C_0\|u_0\|_{ \hdot^{\mu_0}(\mathbb{R}^n) }, 
\ \ 
2C_0AR_{\mu_0}^{p-1}\le 1
\end{equation}
for some constant $C_0>0$.
We obtain the required fixed point by the Banach fixed point theorem.

When $p<p(\mu_0)$, we have $\lim_{S\searrow0} A=0$ since $q(\mu_0)\neq\infty$.
Thus, \eqref{Condition-Contraction} is satisfied for any $u_0\in H^{\mu_0}$ 
and sufficiently small $S=S(\|u_0\|_{\hdot^{\mu_0}(\mathbb{R}^n) })>0$.
When $p= p(\mu_0)$, we have $\lim_{S\searrow0} A=a_0^2$ since $q(\mu_0)=\infty$. 
So that, \eqref{Condition-Contraction} requires the smallness of $\|u_0\|_{\hdot^{\mu_0}(\mathbb{R}^n)}$.
To treat large $\|u_0\|_{\hdot^{\mu_0} (\mathbb{R}^n)}$, we put 
$u_L(s):=U_\pm(s)u_0$
and show that the operator
\begin{equation}
\label{Def-PhiTilde}
\widetilde{\Phi}(u_N)(s):=\pm \int_0^s U_{\pm}(s-\tau) f(\tau, (u_L+u_N)(\tau)) d\tau 
\end{equation}
is a contraction mapping on $X(R_0,R_{\mu_0})$, 
where  
$R_0\ge \max_{j=1,2}\|u_L\|_{ L^{q_j}((0,S), L^{r_j}(\mathbb{R}^n)) }$ 
and 
$R_{\mu_0}\ge \max_{j=1,2}\|u_L\|_{L^{q_j}((0,S), \bdot^{\mu_0}_{r_j2}(\mathbb{R}^n))}$. 
Then the fixed point $u_N$ of $\widetilde{\Phi}$ gives the required solution $u$ by $u=u_L+u_N$.
Similarly to the above argument, we have 
\begin{eqnarray*}
\|\widetilde{\Phi}(u_N)\|_{X^\mu([0,S))}
&\lesssim& \|f(\cdot, u_L+u_N)\|_{L^{q_0'}((0,S), \dot{B}^\mu_{r_0'2}(\mathbb{R}^n))} 
\\
&\lesssim& \frac{\hbar |\lambda|}{m} 
A\|u_L+u_N\|_{ L^{q_1}((0,S), \bdot^{\mu_0}_{r_12}(\mathbb{R}^n)) }^{p-1}
\|u_L+u_N\|_{ L^{q_2}((0,S), \bdot^\mu_{r_22}(\mathbb{R}^n)) }
\\
&\lesssim& \frac{\hbar |\lambda|}{m} A(2R_{\mu_0})^{p-1}2R_\mu
\end{eqnarray*}
for any $\mu$ with $0<\mu\le \mu_0$.
We also have 
\begin{equation*}
\|\widetilde{\Phi}(u_N)\|_{X^0([0,S))}
\lesssim \frac{\hbar |\lambda|}{m}  A(2R_{\mu_0})^{p-1} 2R_0
\end{equation*}
and 
\begin{equation}
\label{Ineq-dPhiTilde}
d(\widetilde{\Phi}(u_N),\widetilde{\Phi}(v_N))
\lesssim \frac{\hbar |\lambda|}{m}  A(2R_{\mu_0})^{p-1} d(u_N,v_N)
\end{equation}
for any $u_N,v_N\in X(R_0,R_{\mu_0})$.
Since $R_{\mu_0}$ can be taken sufficiently small when $S$ is sufficiently small 
by $q_1\neq \infty$ and $q_2\neq \infty$, 
$\widetilde{\Phi}$ is a contraction mapping  
if $S>0$ is sufficiently small.
We note that $S$ depends on the profile of $u_0$ 
in addition to the norm $\|u_0\|_{\hdot^{\mu_0}(\mathbb{R}^n) }$.

The existence time $S$ of the solution $u$ is estimated from below 
by the condition \eqref{Condition-Contraction}.
We recall $1/q(\mu_0)=1-(p-1)(n-2\mu_0)/4$  
and the change of variable 
$s=\int_0^t  a(\tau)^{-2}d\tau$.
When $p=p(\mu_0)$, we have 
\[
A=a_0^2\cdot 
\left\{
\begin{array}{ll}
1& \mbox{if} \ \ a_1\ge0, \\
\left( 1+\frac{a_1n(1+\sigma)T}{2a_0}\right)^{-\frac{8\mu_0}{n(1+\sigma)(n-2\mu_0)}} 
& \mbox{if} \ \ \sigma\neq -1\ \ \mbox{and}\ \ a_1<0, \\
\exp\left({-\frac{4\mu_0 a_1T}{a_0(n-2\mu_0)} }\right)
& \mbox{if} \ \ \sigma= -1\ \ \mbox{and}\ \ a_1<0 \\
\end{array}
\right.
\]
by a direct computation.
So that, the solution $u$ is global for sufficiently small $u_0$ 
if 
(i) $\mu_0=0$, or (ii) $\mu_0>0$ and $a_1\ge0$.
When $p<p(\mu_0)$ and $\sigma\neq -1$, we have 
\[
A^{q(\mu_0)}=
a_0^{(p-1)(n-2\mu_0)q(\mu_0)/2}\cdot
\left\{
\begin{array}{ll}
\frac{2a_0}{n a_1(1-\alpha)(1+\sigma)}
\left(
U^{1-\alpha}-1
\right)
& \mbox{if} \ \ a_1\neq 0 \ \ \mbox{and}\ \ \alpha\neq 1, \\
\frac{2a_0}{n a_1(1+\sigma)}
\log 
U & \mbox{if} \ \ a_1\neq 0 \ \ \mbox{and}\ \ \alpha= 1, \\
T & \mbox{if} \ \ a_1=0, \\
\end{array}
\right.
\]
where we have put 
$U:=1+n a_1 (1+\sigma)T/2a_0$, 
$\alpha:= 2(p-1)\mu_0q(\mu_0)/n(1+\sigma)$.
We note that 
$\alpha =1$ holds if and only if $p=p_1(\mu_0)$,
$\mu_0\neq0$ 
and $\sigma>-1$ hold.
So that, the solution $u$ is global for sufficiently small $u_0$ if 
(iii) $1<p<p(\mu_0)$, $a_1>0$, $\sigma<-1$, or 
(iv) $1<p<p_1(\mu_0)$, $a_1<0$, $\sigma>-1$, or 
(v) $p_1(\mu_0)<p<p(\mu_0)$, $a_1>0$, $\sigma>-1$.
When $p<p(\mu_0)$ and $\sigma= -1$, we have 
\[
A^{q(\mu_0)}=
a_0^{(p-1)(n-2\mu_0)q(\mu_0)/2} 
\cdot
\left\{
\begin{array}{ll}
\frac{1-\exp(-\beta T)}{\beta} 
& \mbox{if}\ \ \beta\neq0, \\
T
& \mbox{if}\ \ \beta=0,
\end{array}
\right.
\]
where we have put $\beta:=a_1(p-1)\mu_0q(\mu_0)/a_0$.
So that, the solution $u$ is global for sufficiently small $u_0$ 
if $\beta> 0$, 
where we note that $\beta> 0$ holds if and only if $a_1>0$, $p>1$ 
and $\mu_0>0$ hold.


We prove the continuous dependence of solutions on initial data.
Let $\{u_{j0}\}_{j\ge1}$ be a sequence of data which converges to $u_0$ 
in $H^{\mu_0}(\mathbb{R}^n)$, 
and let $\{u_j\}_{j\ge1}$ be the solutions of \eqref{Cauchy} for $\{u_{j0}\}_{j\ge1}$.
We use the equation 
\[
(u-u_j)(s)=U_\pm (s)(u_{0}-u_{j0}) \pm  \int_0^s U_{\pm}(s-\tau) 
\left\{ f(\tau,u(\tau)) - f(\tau,u_j(\tau)) \right\} d\tau
\]
for $j\ge1$, and the similar argument for \eqref{Es-Metric} to obtain 
\begin{equation}
\label{Ineq-Power-d}
d(u,u_j)\le C_0\|u_0-u_{j0}\|_{L^2(\mathbb{R}^n)} 
+
C_0A\max_{w=u,u_j} \|w\|_{L^{q_1} ((0,S), \bdot^{\mu_0}_{r_12}(\mathbb{R}^n)) }^{p-1}d(u,u_j) 
\end{equation}
for some constant $C_0>0$.
So that, we have $d(u,u_j)\rightarrow0$ 
as $j\rightarrow\infty$ 
since $\lim_{S\searrow0}A=0$ when $p<p(\mu_0)$.
When $p=p(\mu_0)$, we use the argument starting from \eqref{Def-PhiTilde}.  
The nonlinear part $u_{jN}$ of $u_j$ is obtained with the property 
\begin{equation}
\label{Ineq-CD-ujN}
\max_{k=1,2} \|u_{jN}\|_{L^{q_k} ((0,S), \bdot^{\mu_0}_{r_k2}(\mathbb{R}^n)) }
\le \max_{k=1,2} \|u_{jL}\|_{L^{q_k} ((0,S), \bdot^{\mu_0}_{r_k2}(\mathbb{R}^n)) }
\end{equation}
under the condition 
\[
2C_1 A
\left(2\max_{k=1,2}\|u_{jL}\|_{L^{q_k}((0,S), \bdot^{\mu_0}_{r_k2}(\mathbb{R}^n))  }
\right)^{p-1}\le 1
\]
for some constant $C_1>0$, 
where $u_{jL}:=U_\pm(\cdot)u_{j0}$.
We are able to take $S>0$ uniformly for the existence time of $\{u_j\}_{j=1}^\infty$ since 
\begin{multline}
\label{Ineq-CD-ujL}
\max_{k=1,2}\|u_{jL}\|_{ L^{q_k}((0,S), \bdot^{\mu_0}_{r_k2}(\mathbb{R}^n))  }
\le 
\max_{k=1,2}\|u_{L}\|_{L^{q_k}((0,S), \bdot^{\mu_0}_{r_k2}(\mathbb{R}^n)) }
\\
+C\|u_0-u_{j0}\|_{\hdot^{\mu_0}(\mathbb{R}^n) }
\end{multline}
by the energy and Strichartz estimates,  
and $u_{j0}\rightarrow u_0$ as $j\rightarrow\infty$.
By \eqref{Ineq-CD-ujN} and \eqref{Ineq-CD-ujL}, we have 
\[
\max_{k=1,2}\|u_j\|_{L^{q_k}((0,S),\bdot^{\mu_0}_{r_k2}(\mathbb{R}^n)) }
\le 2\max_{k=1,2} \|u_L\|_{L^{q_k}((0,S),\bdot^{\mu_0}_{r_k2}(\mathbb{R}^n))}
+
2C\|u_0-u_{j0}\|_{\hdot^{\mu_0}(\mathbb{R}^n) }.
\]
Since the right hand side tends to zero as $j\rightarrow\infty$ and $S\searrow0$ 
by $q_1\neq\infty$ and $q_2\neq\infty$, 
we obtain 
$d(u,u_j)\rightarrow0$ 
as $j\rightarrow\infty$ 
by \eqref{Ineq-Power-d}.

%

\newsection{Proofs of Corollaries}
Let $V_1(z)$ be a complex-valued function for $z\in \mathbb{C}$.
We consider the conservation law for the equation 
\begin{equation}
\label{Eq-W1}
\pm i\frac{2m}{\hbar}\partial_s u(s,x)
+e^{-2i\omega}\,\Delta u(s,x)
- 
e^{-2i\omega} \frac{a^2}{w}(s) V_1(uw)(s,x)=0
\end{equation}
for $0\le s<S$ and $x\in \mathbb{R}^n$.

\begin{lemma}
\label{Lemma-Conservation}
Let $V_1$ satisfy $\mathrm{Im}\{\bar{z} V_1(z)\}=0$ for any $z$ and its complex conjugate $\bar{z}$.
For the solution $u$ of  \eqref{Eq-W1}, the following results hold.

(1) 
The charge of $u$ satisfies 
\begin{multline*}
\frac{m}{\hbar} \int_{\mathbb{R}^n} |u(s,x)|^2 dx
\pm
\sin 2\omega \int_0^s \int_{\mathbb{R}^n} |\nabla u(\tau,x)|^2
+
\frac{a^2}{w}(\tau) (\bar{u}V_1(uw)) (\tau,x)dx d\tau
\\
=
\frac{m}{\hbar} \int_{\mathbb{R}^n} |u(0,x)|^2 dx.
\end{multline*}

(2) 
If there exists a function $V_0$ which satisfies  $\partial_s\{V_0(v)\}=\mathrm{Re}\{(\partial_s \bar{v}) V_1(v)\}$ 
for any function $v$, 
then the energy of $u$ satisfies  
\begin{equation}
\label{Eus-Eu0}
E(u)(s)\pm \frac{2m\sin 2\omega}{\hbar}
\int_0^s\int_{\mathbb{R}^n} |\partial_s u(\tau,x)|^2 dx d\tau
+
\int_0^s\int_{\mathbb{R}^n} I(\tau,x) dx d\tau
=E(u)(0),
\end{equation}
where 
\begin{equation}
\label{Conservation-Energy}
E(u)(s):=\int_{\mathbb{R}^n} \frac{1}{2}|\nabla u(s,x)|^2
+
\frac{a^2}{w^2}
(s)V_0(uw)(s,x)dx
\end{equation}
and 
\[
I(\tau,x):=
\frac{n}{2a_0^n} a(\tau)^{n+1} 
\frac{da}{ds}(\tau) 
\left(
\bar{u} w V_1(uw)-\frac{2(n+2)}{n} V_0(uw)
\right)(\tau,x).
\]

\begin{proof}
(1) Multiplying $\bar{u}$ to the equation \eqref{Eq-W1} 
and taking its imaginary part, we obtain 
\begin{equation}
\label{Charge-Diff}
\pm \frac{m}{\hbar} \partial_s |u|^2
+
\nabla \cdot \mbox{\rm Im}
\left(
e^{-2i\omega} \bar{u}\nabla u
\right)
+
\left(
|\nabla u|^2
+
\frac{a^2}{w} \bar{u} V_1(uw)
\right)
\sin 2\omega=0.
\end{equation}
We obtain the required result by the integration for $s$ and $x$.

(2) Multiplying $\partial_s\bar{u}$ to \eqref{Eq-W1} and taking its real part, we have
\[
\partial_s e(u)
-
\nabla\cdot \mbox{\rm Re} (\partial_s\bar{u} \nabla u)
\pm
\frac{2m\sin 2\omega}{\hbar} |\partial_s u|^2
+J=0,
\]
where we have put 
$e(u):=|\nabla u|^2/2+a^2w^{-2} V_0(uw)$ and 
\[
J:=
\frac{1}{2}
\left\{
a^2\partial_s (w^{-2})
\bar{u}wV_1(uw)
-
2\partial_s(a^2 w^{-2})V_0(uw) 
\right\}.
\]
We note  $I=J$ since $w=(a_0/a)^{n/2}$.
We obtain the required result by the integration for $s$ and $x$.
\end{proof}
\end{lemma}


We put 
$V_0(v):=\lambda  |v|^{p+1}/(p+1)$ and $V_1(v):=\lambda  |v|^{p-1}v$ for  
$\lambda \in \mathbb{R}$.
Then $V_0$ and $V_1$ satisfy the assumption in 
Lemma \ref{Lemma-Conservation}, 
and $I$ in the lemma is rewritten as 
\begin{equation}
\label{I-Rewrite}
I(\tau,x)=
\frac{n}{2a_0^n} a(\tau)^{n+1} 
\frac{da}{ds}(\tau) 
V_0(uw)(\tau,x)\left(p-1-\frac{4}{n}\right).
\end{equation}


Let us prove Corollary \ref{Cor-GlobalEnergy-Power}.
Since the existence time of the solutions 
obtained by Theorem \ref{Th-Power-LocalGlobal}
is dependent only on the norms of the data when $p<p(\mu_0)$, 
it suffices to show the uniform bound of the solutions in 
$L^2(\mathbb{R}^n)$ when $\mu_0=0$, 
and $H^1(\mathbb{R}^n)$ when $\mu_0=1$. 
First, we show the case $\mu_0=0$.
For any fixed $u_0\in L^2(\mathbb{R}^n)$, 
let $\{u_{j0}\}_{j\ge1} \subset H^1(\mathbb{R}^n)$ be a sequence of data which converges to $u_0$ in $L^2(\mathbb{R}^n)$, 
and let $\{u_j\}_{j\ge1}$ be the solutions of \eqref{Cauchy} for $\{u_{j0}\}_{j\ge1}$ 
which are obtained by Theorem \ref{Th-Power-LocalGlobal}.
By the continuous dependence of the solutions on the data, 
there exists $S$ with $0<S\le S_0$ such that 
$u_j$ converges to $u$ in 
$L^\infty((0,S),L^2(\mathbb{R}^n))
\cap 
\bigcap_{k=1}^2 L^{q_k}((0,S),L^{r_k}(\mathbb{R}^n))$ as $j$ tends to infinity.
By (1) in Lemma \ref{Lemma-Conservation}, $0\le \pm \omega\le \pi/2$  
and $\lambda \ge0$, we have 
$\int_{\mathbb{R}^n} |u_j(s,x)|^2 dx \le \int_{\mathbb{R}^n} |u_j(0,x)|^2 dx$.
Taking the limit of $j$, we have the required bound  
$\|u\|_{L^\infty((0,S),L^2(\mathbb{R}^n))}\le \|u_0\|_{L^2(\mathbb{R}^n)}$. 
Next, we show the case $\mu_0=1$.
By \eqref{I-Rewrite}, we have $I\ge0$ if and only if 
$\lambda  a_1(p-1-4/n)\ge0$.
Thus, we have 
$\|\nabla u(s,\cdot)\|_{L^2(\mathbb{R}^n)}^2\le 2E(u)(s)\le 2E(u)(0)$ 
for $0\le s<S$ by (2) in Lemma \ref{Lemma-Conservation}, $\lambda>0$, 
$a_1(p-1-4/n)\ge0$ and $0\le\pm \omega\le\pi/2$.
By (1) in Lemma \ref{Lemma-Conservation}, 
we also have 
$\|u(s,\cdot)\|_{L^2(\mathbb{R}^n)}\le \|u(0,\cdot)\|_{L^2(\mathbb{R}^n)}$ 
for $0\le s<S$.
So that,   
$\|u\|_{L^\infty((0,S),H^1(\mathbb{R}^n))}$ is uniformly bounded 
independent of $S$.


Let us prove Corollary \ref{Cor-GlobalEnergy-Power-Focusing}.
By \eqref{I-Rewrite} and $\lambda a_1(p-1-4/n)\ge0$, 
we have $E(u)(s)\le E(u)(0)$ for $s\ge0$.
By the Gagliardo-Nirenberg inequality 
\[
\|u\|_{L^{p+1}(\mathbb{R}^n)}\lesssim \|u\|_{L^2(\mathbb{R}^n)}^{1-n(p-1)/2(p+1)}\|\nabla u\|_{L^2(\mathbb{R}^n)}^{n(p-1)/2(p+1)},
\] 
we have 
\begin{multline}
\label{EuDu}
E(u)(s)\ge 
\left\{
1-\frac{2C|\lambda|(a^2w^{p-1})(s) }{p+1}\|u(s)\|_{L^2(\mathbb{R}^n)}^{p+1-n(p-1)/2}
\|\nabla u(s)\|_{L^2(\mathbb{R}^n)}^{n(p-1)/2-2}
\right\}
\\
\cdot
\frac{\|\nabla u(s)\|^2_{L^2(\mathbb{R}^n)} }{2}
\end{multline}
for some constant $C>0$.
By $\omega=0$ or $\omega=\pi/2$, the conservation of charge 
$\|u\|_{L^2(\mathbb{R}^n)}=\|u_0\|_{L^2(\mathbb{R}^n)}$,
and the condition $p< 1+4/n$, 
the inequality \eqref{EuDu} shows that 
$\|\nabla u\|_{L^2(\mathbb{R}^n)}$ does not blow up before $S_0$.
Namely, $u$ is a global solution.


Let us prove Corollary \ref{Cor-GlobalEnergy-Power-Focusing-BlowUp-Parabolic}.
By \eqref{I-Rewrite} and $\lambda a_1(p-1-4/n)\ge0$, we have $I\ge0$.
Thus, by (2) in Lemma \ref{Lemma-Conservation}, we have 
\begin{equation}
\label{PartialE}
E(u)(s)\le 
E(u)(0) 
\mp \frac{2m\sin2\omega}{\hbar} 
\int_0^s\int_{\mathbb{R}^n} |\partial_s u(\tau,x)|^2 dxd\tau.
\end{equation}
On the other hand, by \eqref{Charge-Diff} and 
$\overline{uw} V_1(uw)=(p+1)V_0(uw)$, we have 
\begin{equation}
\label{PartialU}
\partial_s \int_{\mathbb{R}^n} |u(s,x)|^2 dx \ge \mp \frac{\hbar\sin 2\omega}{m}  (p+1)E(u)(s).
\end{equation}
We put $K(s):=\int_0^s \int_{\mathbb{R}^n} |u(\tau,x)|^2 dxd\tau +K_0$ 
for a constant $K_0>0$ which is determined later.
By \eqref{PartialE} and \eqref{PartialU}, we have 
\[
\partial_s^2 K(s)\ge \mp \frac{\hbar\sin2\omega}{m} (p+1)
\left\{
E(u)(0)\mp \frac{2m\sin2\omega}{\hbar}
\int_0^s\int_{\mathbb{R}^n} |\partial_s u(\tau,x)|^2 dxd\tau
\right\}.
\]
Since $\partial_sK(s)$ satisfies 
\[
\partial_s K(s)=\int_{\mathbb{R}^n} |u_0(x)|^2 dx
+
2\mbox{\rm Re} \int_0^s\int_{\mathbb{R}^n} (u\partial_s \bar{u}) (\tau,x)dxd\tau
\]
by the definition of $K(s)$, 
we have 
\begin{multline*}
\left(\partial_s K(s)\right)^2
\le
\left(1+\frac{1}{\varepsilon}\right)
\left(
\int_{\mathbb{R}^n} |u_0(x)|^2 dx 
\right)^2
\\
+
4(1+\varepsilon)
\int_0^s \int_{\mathbb{R}^n} |u(\tau,x)|^2 dxd\tau 
\int_0^s \int_{\mathbb{R}^n} |\partial_s u(\tau,x)|^2 dxd\tau 
\end{multline*}
for any $\varepsilon>0$ by the H\"older inequality.
So that, for any $\alpha$ with $\alpha>-1$, we have 
\begin{multline*}
\partial_s^2 K(s) K(s)-(1+\alpha)\left(\partial_s K(s)\right)^2
\ge 
\mp \frac{\hbar\sin2\omega}{m}(p+1)
E(u)(0)K_0
\\
-
(1+\alpha)\left(1+\frac{1}{\varepsilon}\right)
\left(\int_{\mathbb{R}^n} |u_0(x)|^2 dx\right)^2
+
2\left\{
(\sin2\omega)^2(p+1)-2(1+\alpha)(1+\varepsilon)
\right\}
\\
\cdot
\int_0^s \int_{\mathbb{R}^n} |u(\tau,x)|^2 dxd\tau 
\int_0^s\int_{\mathbb{R}^n} |\partial_s u(\tau,x)|^2 dxd\tau.
\end{multline*}
Therefore, if $\omega\neq0$, $\omega\neq \pi/2$, 
$E(u)(0)<0$, $K_0$ is sufficiently large, 
$\alpha>0$ and $\varepsilon>0$ 
are sufficiently small, and $p>2/(\sin 2\omega)^2-1$, 
then  we have 
$\partial_s^2 K(s) K(s)-(1+\alpha)\left(\partial_s K(s)\right)^2>0$.  
So that, we have 
\[
\frac{1}{K(s)^{\alpha}}
<
\frac{K(0)-s\alpha\|u_0\|_{L^2(\mathbb{R}^n)}^2}{ K(0)^{1+\alpha} }.
\]
since 
$\partial_s( \partial_s K/K^{1+\alpha})
=(\partial_s^2KK-(1+\alpha)(\partial_sK)^2)/K^{2+\alpha}>0$.
This inequality shows $K(s)$ blows up at finite time 
$s=S_1:=K(0)/\alpha\|u_0\|_{L^2(\mathbb{R}^n)}^2$, 
which shows that there exists a sequence $\{s_j\}_{j\ge1}$ 
which converges to $S_1$ such that 
$\int_{\mathbb{R}^n} |u(s_j, x)|^2 dx$ tends to infinity as $j\rightarrow\infty$.


Let us prove Corollary 
\ref{Cor-GlobalEnergy-Power-Focusing-BlowUp-Dispersive}.
We use the virial identities 
$\partial_s \int_{\mathbb{R}^n} |x|^2|u(s,x)|^2dx
=\pm 2\hbar e^{-2i\omega}{\rm Im} \int_{\mathbb{R}^n} \bar{u} x\cdot\nabla u dx/m$, and 
\begin{multline*}
\partial_s^2 \int_{\mathbb{R}^n} |x|^2|u(s,x)|^2dx
\\
=
\left(
\frac{\hbar}{m}
\right)^2
\left\{
2\int_{\mathbb{R}^n} |\nabla u(s,x)|^2 dx
+\frac{n(p-1) a(s)^2}{w(s)^2}
\int_{\mathbb{R}^n} 
V_0(uw)(s,x)dx
\right\},
\end{multline*}
which are derived from the first equation in \eqref{Cauchy}.
Since we have 
\[
\frac{a(s)^2}{w(s)^2}
\int_{\mathbb{R}^n} 
V_0(uw)(s,x)dx
=E(u)(s)-\frac{1}{2}\int_{\mathbb{R}^n} |\nabla u(s,x)|^2 dx 
\]
by the definition of $E(u)$,
we have 
\begin{multline*}
\partial_s^2 \int_{\mathbb{R}^n} |x|^2|u(s,x)|^2dx
\\
\le 
\left(
\frac{\hbar}{m}
\right)^2
\left\{
n(p-1)E(u)(s)-\frac{n}{2}\left(p-1-\frac{4}{n}\right)
\int_{\mathbb{R}^n} |\nabla u(s,x)|^2 dx
\right\}.
\end{multline*}
So that, we have 
\begin{equation}
\label{Ineq-Ds2}
\partial_s^2 \int_{\mathbb{R}^n} |x|^2|u(s,x)|^2dx
\le 
\frac{\hbar^2 n(p-1)}{m^2}
E(u)(s)
\end{equation}
if $p\ge 1+4/n$.
Since $E(u)(s)\le E(u)(0)$ by $a_1\lambda (p-1-4/n)\ge0$, 
the left hand side in \eqref{Ineq-Ds2} is bounded by a negative number from above 
if $E(u)(0)<0$.
Therefore, there exists $S_2>0$ such that 
$\lim_{s\nearrow S_2}\int_{\mathbb{R}^n} |x|^2|u(s,x)|^2dx=0$.
By the Heisenberg uncertainty principle 
\begin{equation}
\|u\|_{L^2(\mathbb{R}^n)}^2\le \frac{2}{n} \|\nabla u\|_{L^2(\mathbb{R}^n)}\||x|u\|_{L^2(\mathbb{R}^n)}
\end{equation}
and the conservation of charge 
$\|u(\cdot)\|_{L^2(\mathbb{R}^n)}=\|u_0\|_{L^2(\mathbb{R}^n)}$,  
$\|\nabla u(s,\cdot)\|_{L^2(\mathbb{R}^n)}$ tends to infinity as $s$ tends to $S_2$.
So that, the solution $u$ blows up in finite time.

%


%

\end{document}